\documentclass[12pt]{article}\usepackage[hyperfootnotes=false]{hyperref}
\usepackage{epsfig}
\usepackage{float}
\usepackage{empheq}
\usepackage{bbold}

\usepackage{xspace}

\makeatletter

\@addtoreset{equation}{section}
\makeatother

\usepackage[utf8]{inputenc}
\usepackage{amsmath}
\usepackage{caption}
\usepackage{amsmath}
\usepackage{amssymb}
\usepackage{graphicx}

\newlength{\abstractwidth}
\renewcommand{\thefootnote}{\fnsymbol{footnote}}
\renewcommand{\thanks}[1]{\footnote{#1}} 
\newcommand{\starttext}{
\setcounter{footnote}{0}
\renewcommand{\thefootnote}{\arabic{footnote}}}

\newcommand{\be}{\begin{equation}}
\newcommand{\bea}{\begin{eqnarray}}
\newcommand{\eea}{\end{eqnarray}}
\newcommand{\beq}{\begin{equation}}
\newcommand{\ee}{\end{equation}}

	\newcommand*\widefbox[1]{\fbox{\hspace{2em}#1\hspace{2em}}}

	\def\dsp.{de Sitter space.}
	\def\eq{&=&}

	\def\ra{\rangle}
	\def\simleq{\; \raise0.3ex\hbox{$<$\kern-0.75em
			\raise-1.1ex\hbox{$\sim$}}\; }
	\def\simgeq{\; \raise0.3ex\hbox{$>$\kern-0.75em
			\raise-1.1ex\hbox{$\sim$}}\; }
	
	\def\bi{\begin{itemize}}
		\def\ei{\end{itemize}}

	\def\dof{degrees of freedom }
	
	\def\CA{{\cal{A}}}

	\def\CF{{\cal{F}}}

	\def\CJ{{\cal{J}}}

	\def\CO{{\cal{O}}}

	\def\Tr{\rm Tr \it}
	\def\bsub{ \begin{subequations}
			\begin{empheq}[box=\widefbox]{align}  }
			\def\esub{ \end{empheq}
	\end{subequations}}
	
	\def\1{\(  \mathbb{1} \)}

	\def\lf{\left(}
	\def\rg{\right)}

	\def\bn{\bigskip \noindent}

	\def\dk{${\rm DSSYK_{\infty}}$}

        \def\g{g_{ym}}

\def\w{\langle W \rangle}  
\def\W{\big( \langle W \rangle \big) }

	\makeatletter
	\g@addto@macro\normalsize{%
		\setlength\abovedisplayskip{10pt}
		\setlength\belowdisplayskip{20pt}
		\setlength\abovedisplayshortskip{10pt}
		\setlength\belowdisplayshortskip{20pt}
	}
	\makeatother
	
	\usepackage{color}

	\usepackage{authblk}
	\title{\Large \bf DSSYK at Infinite Temperature: The Flat-Space Limit and the 't~Hooft Model}

	\author[1]{\Large Shoichiro Miyashita}
	\author[2]{\Large Yasuhiro Sekino}
	\author[3,4]{\Large Leonard Susskind}
	\affil[1]{Department of Physics, National Dong Hwa University,
Hualien 97401, Taiwan, R.O.C. \vspace{1em}}
	\affil[2]{Department of Liberal Arts and Sciences,
Faculty of Engineering, Takushoku University, 
Hachioji, Tokyo 193-0985, Japan \vspace{1em}}
	\affil[3]{Stanford Institute for Theoretical Physics and Department of Physics, Stanford University, Stanford, CA 94305-4060, USA \vspace{1em}}
	\affil[4]{Google, Mountain View, CA, USA}
	\date{}
	\usepackage{upgreek}

\begin{document}
		
		\begin{titlepage}
			\maketitle
			
			\begin{abstract}
				\Large
In the limit of infinite radius de Sitter space becomes locally flat and the static patch tends to Rindler space.  A holographic  description of the static patch  must result in a holographic  description of some flat space theory, expressed in Rindler coordinates. Given such a holographic theory how does one decode the hologram and determine the bulk flat space theory, its particle spectrum, forces, and bulk  quantum fields? In this paper we will answer this question  for a particular case: DSSYK at infinite temperature  and show that the bulk theory is a strongly coupled version of the 't~Hooft model, i.e., (1+1)-dimensional QCD, with a single quark flavor. It may also be thought of as  an open string theory with mesons lying on a single Regge trajectory.

\end{abstract}

		\end{titlepage}
		
		\rightline{}
		\bigskip
		\bigskip\bigskip\bigskip\bigskip
		\bigskip
		
		\starttext \baselineskip=17.63pt \setcounter{footnote}{0}

		\LARGE
		
		\tableofcontents
		
\section{The Holographic Principle}\label{HP}

A naive version of the Holographic Principle states that a region of space $\Gamma$ and everything in it can be described by a quantum system with a Hilbert space of dimension $\exp{A/4G},$ $A$ being the area of the boundary of $\Gamma$; and a Hamiltonian acting on that space
 \cite{tHooft:1993dmi}\cite{Susskind:1994vu}.  A more precise version  \cite{Bousso:1999cb} goes as follows:

Consider a nested family of regions $\Gamma(s)$ parameterized by a continuous variable $s$. For small $s$ the area $A(s)$ will grow monotonically. $A(s)$ may grow forever and approach infinity or it may reach a maximum at some value of $s$ and then decrease back to zero. (Other behaviors are possibe but we will consider only these two.) One defines a holographic screen to be the surface of maximum area as $s$ varies. If $A(s)\to \infty$ then the screen is at infinity. If the spatial slice is a sphere  the maximum area is finite. For the case of 2-dimensional space it is a great circle.  The Holographic principle says that everything in the interior of the screen may be described  by a theory with a Hamiltonian acting on a Hilbert space of dimension $\exp{A/4G}.$ The first case with a screen at infinity is exemplefied by AdS; the second case with finite screen, by de Sitter space. In the latter case the screen is the (stretched) horizon and the region interior to  the screen is the static patch.
       
In addition to a Hamiltonian and a Hilbert space one must  specify a density matrix for the static patch. Following 
\cite{Dong:2018cuv}\cite{Chandrasekaran:2022cip} we assume the density matrix is maximally mixed, or what is equivalent, the Boltzmann 
temperature \cite{Rahman:2024vyg}	 is infinite,
\be 
T_{B} = \infty.
\label{Tb}
\ee
The usual Hawking temperature is a derived or emergent quantity  called Tomperature in  \cite{Lin:2022nss}.

\subsection{The Flat-Space Limit} \label{fsl}
The flat-space limit of AdS/CFT is well known 
\cite{Polchinski:1999ry}\cite{Susskind:1998vk}. It is the limit $N\to \infty$ with the CFT gauge coupling $\g$ held fixed\footnote{This limit is not the standard 't~Hooft limit,  $\g^2N$ held fixed.}, and energies scaling like $N^{1/4}$ in units of the AdS radius. The corresponding flat-space limit of de Sitter space is similar. It is 
 the limit of large entropy in which the de Sitter radius of curvature goes to infinity and spacetime becomes  locally flat. The static patch, out to arbitrary distances from the horizon, becomes flat Rindler space. 
 
 The question addressed in this paper is: Given a holographic setup of the kind described above  how do we discover the dual bulk theory that it describes: the particle spectrum; forces; fields; and interactions? No attempt will be made to answer this question in generality. We will  consider one specific example, \dk, the holographic dual of JT-gravity with positive cosmological constant and a matter sector to be determined. In the flat--space limit the theory must be dual to some quantum field theory expressed in Rindler coordinates.

\section{DSSYK$_{\infty}$} \label{DSSYK}

For definiteness we will consider complex SYK with $N$ species of complex fermions $\bar{\psi}_i,\ \psi^i.$ The Hilbert space is $2^N$-dimensional and at infinite Boltzmann temperature the entropy is exactly,
\be 
S =N \log{2}.
\label{S=NLg2}
\ee 

The Hamiltonian in string units
is 
\be 
H = \sum J^{j_1j_{2}\cdots j_p}_{i_1i_{2}\cdots i_p} \bar{\psi}_{j_1}\bar{\psi}_{j_2}\cdots\bar{\psi}_{j_p}    \psi^{i_p}\cdots \psi^{i_1}.
\label{H}
\ee
The sum in \eqref{H} is over all $1\le i_1<i_2\cdots\le N$ and $1\le j_1<j_2\cdots\le N.$ The coupling constants $J^{j_1j_2\cdots j_p}_{i_1i_2\cdots i_p},$ which will collectively be donoted $\{ J \},$ are independently drawn from a Gaussian random ensemble 
with variance 
\be 
\text{Var}(J) = \CJ^2 \  \frac{N}{p^2} \  \frac{p! p!}{N^pN^p}
\label{var}
\ee
where $\CJ$ is a numerical constant with units of energy. We note that the scaling in \eqref{var} is string-scale, rather than cosmic scale:
$\CJ$ in this paper is defined to be $p$ times the constant $\CJ$ used in previous papers such as \cite{Susskind:2022bia, Sekino:2025bsc}, and is interpreted as the inverse of the string length.

The double-scaled limit is defined by,
\bea  
N&\to& \infty \cr \cr
\frac{p^2}{N} &\equiv& \lambda =\text{fixed}.
\label{DSlim}
\eea
The constant $\lambda$ is the only adjustable  dimensionless parameter in \dk. The Boltzmann temperature is not adjustable but is infinite  \eqref{Tb}.

The flat-space limit is simply the $N\to \infty$  limit of \dk \ with the additional stipulation that energies and time-scales are kept fixed as $N\to \infty.$

\section{Expectation Values and Ensemble Averages} \label{expav}
Operators in \dk \ will depend on the fermionic variables $\psi, \ \bar{\psi}$ and also on the couplings $\{ J  \}.$ To see why they may depend on $\{ J  \}$ consider a Heisenberg operator 
$$\psi(t) = e^{-iHt}\psi e^{iHt}.$$ 
It will depend on $\{ J  \}$ through the dependence on the Hamiltonian $H.$ The exponentials $e^{\pm i Ht}$ may be expanded in powers of $\CJ$ giving rise to a chord expansion \cite{Berkooz:2018jqr}.

A typical example of operators that have a non-trivial dependence on $\{J\}$ are products of fields at different times such as $$\bar{\psi}(0) \psi(t).$$

A general operator which depends on both the fermions and the couplings will be denoted $\CO(\psi,  J).$ The  expectation value of $\CO$ is given by,
\be 
\langle \CO \rangle = \Tr \  \CO(\psi, J)
\label{<o>}
\ee
where $\Tr$ denotes the normalized trace  over the fermion Hilbert space, defined so that the trace of the identity is one. The expectation value will in general depend on $\{J\}.$

The ensemble average of a function  $F(J)$ is defined as an integral over $J$ weighted by the gaussian measure with variance \eqref{var}. We will denote it by large round brackets\footnote{%
In the previous paper~\cite{Sekino:2025bsc} by two of the present authors,
ensemble average over $J$ was denoted by $\langle \ \rangle_J$.}%
, i.e., 
\bea
\Big( F \Big) &=& \int P(J) F(J) dJ \cr \cr
P(J) &\equiv& e^{- \sum \frac{J^2}{\text{Var} (J)}} \cr \cr
J^2  &\equiv&  \left| J^{j_1j_2\cdots j_p}_{i_1i_2\cdots i_p} \right|^2
\label{average}
\eea
and Var given in \eqref{var}.

SYK was originally set up to calculate
ensemble averages  of  expectation values,
\be 
\Big( 
\langle \CO  \rangle  \Big)=  \int  dJ  \ P(J) \ \Tr \CO.
\label{avexp}
\ee

\subsection{Self-Averaging} \label{selfav}

The dependence of $\langle \CO  \ra$ on $\{J\} $ is a measure of how much variation there is, not in the measured value of $\CO,$ but in the expectation value of $\CO$ as we vary over the ensemble. The variance in the expectation value is defined by,
\be  
\text{Var}\langle \CO  \rangle = 
\big
( \langle \CO  \rangle ^2 
\big)
-  \big
( \langle \CO  \rangle 
\big)^2.
\label{varO}
\ee
In general there is no reason to expect $\text{Var}\langle \CO  \rangle $ to be very small, and therefore no reason for the average of the expectation value to be a good approximation to the expectation value itself. But if $\text{Var}\langle \CO  \rangle $ is very small then we say that $\CO$ is self-averaging. For example we may find that for some reason $$\text{Var}\langle \CO  \rangle \sim N^{-\alpha}$$ for some $\alpha > 0.$ In that case $\CO$ is self-averaging to order $\alpha.$ One may even find $\text{Var}\langle \CO  \rangle$ tends to zero faster than any power of $N.$ Then   $\CO$ is ``perfectly self-averaging." We will see later in the next section that in the double-scaled limit many important quantities are perfectly self-averaging.

\section{Decoding the Hologram}

By decoding a hologram we mean determining the bulk theory from the boundary holographic theory. In AdS/CFT  conformal symmetry and supersymmetry go a long way toward this goal, but in \dk \ (or very likely in any holographic theory of de Sitter space) these special features are absent and we will have to rely on other ``tricks."  One fairly general method is to calculate \dk \ correlation functions and attempt to interpret them in  terms of correlations on the stretched horizon of de Sitter space. This, if successful, would be highly persuasive but the simple truth is that we have not made much progress along these lines. One obstruction explained in 	\cite{Susskind:2023hnj}  is that the degrees of freedom that can propagate into the bulk of the static patch are a vanishingly small subset of the full set that comprise the enormous entropy of de Sitter space. One faces the ``needle in the haystack" problem of isolating those degrees of freedom before we even start to calculate their matrix elements.

The other time-honored method is to avoid the difficulties of dynamics and rely instead on symmetries. This is the path we will take here. Once the symmetries are well-understood the dynamics may prove easier to understand.

\subsection{Symmetries of DSSYK}
What are the symmetries of  \dk? First of all there are discrete symmetries such as time-reversal and in the complex  case, charge conjugation. Also in the complex case there is a $U(1)$ global symmetry associated with charge conservation,
\bea 
\psi^i &\to& e^{i\theta} \psi^i \cr \cr
\bar{\psi}_i &\to & e^{-i\theta} \bar{\psi}_i.
\label{U1}
\eea
The charge associated with the $U(1)$ symmetry is given by,
\be 
Q=\sum_i \bar{\psi}_i\psi^i.
\label{charge}
\ee
The $U(1)$ symmetry, the conservation of charge,  as well as the discrete symmetries are exact and apply for any values of  the coupling constants $\{J\}.$ They are exact for both the ensemble-averaged theory and also for the unaveraged theory, i.e., for each instance of the random choice of $\{J\}.$

The bulk implications of this exact global $U(1)$ symmetry  are well understood: Boundary symmetries become gauge symmetries in the bulk theory. That is a powerful fact which in $(1+1)$-dimensions  tells us that the the actual propagating fields in  the bulk must be $U(1)$ neutral---``electrically" neutral so to speak. All charged degrees of freedom are trapped near the stretched horiozon. We will discuss this more fully later.

\subsection{Symmetries of the Averaged Theory}

Usually when one speaks of the SYK theory one is referring to the ensemble averaged theory defined by integrating over $\{J\}$ as in \eqref{avexp}. There are symmetries of the averaged theory that are not symmetries of indivisual instances of the ensemble. To distinguish these symmetries we will call them symmetries of the average.

The most important symmetry of the average is $SU(N),$ the symmetry which unitarily transforms the $N$ fermion operators according to,
\bea 
\psi^i \to U^i_j \psi^j \cr \cr
\bar{\psi}_i \to \bar{\psi}_j U^j_i
\label{sun}
\eea
with $U$ belonging to $SU(N).$

It is obvious from the form of $H$ \eqref{H} that \eqref{sun} is not a symmetry
unless the $J^{j_1j_2\cdots j_p}_{i_1i_2\cdots i_p}$ coefficients are treated as tensors which are allowed to transform. Why is this allowed? The answer is that it is not allowed for the unaveraged theory. But for the averaged theory in which the $\{J\}$ are integrated over with the $SU(N)$ invariant measure in \eqref{average}, it is allowed. Correlation functions in the ensemble averaged theory are $SU(N)$ invariant. As emphasized above, this in itself is no guarantee that the symmetry is accurate, even approximately, for the unaveraged theory.

\subsection{Diagnostics}
It will be useful to have a set of correlation functions whose nonvanshing 
would signal a breaking of $SU(N).$ If all such symmetry-breaking diagnostics vanish we may conclude that $SU(N)$ is a symmetry. One way to construct such diagnostic operators is to begin with two fermionic operators which transform differently. An example would be,
\bea 
S &=&\bar{\psi}_i\psi^i  \ \ \ \ \  \text{(sum implicit)}  \cr  \cr
A \eq \ \bar{\psi}_i\psi^j  \ \ \ \ \  (i\neq j)
\label{sngadj}
\eea
$S$ is a singlet and $A$ transforms under the adjoint of $SU(N).$ The equal time correlation function $$\Tr (S A)$$
vanishes but  
\be 
C(t, J)=  \langle S(0) A(t) \rangle =Tr S e^{-iHt}Ae^{iHt}
\label{C(t)   }
\ee
in general will not vanish since $H$ is not invariant under $SU(N)$. A non-zero value of $C(t,J)$ would be an unambiguous signal that $SU(N)$ is not a symmetry of the unaveraged theory. There is every reason to expect this to be the  case.

By contrast the ensemble-averaged counterpart,
\be  
\big(  C(t, J) \big) =     \int   P(J) dJ  \big\{Tr S e^{-iHt}Ae^{iHt}        \big\}
\ee
 must vanish because the integration over $J$ with the invariant measure $P(J)dJ$ is $SU(N)$ invariant.

There are an infinite number of operators of the general form \eqref{C(t)   } which can serve as diagnostics, we will denote them by  $W$. To construct them the singlet $S$ can be replaced by any other operator that transformes differently from $A$ and the time can be any value. We will assume that the vanishing of all such averaged expectation values $\W$  is proof that the averaged  theory is invariant under $SU(N).$

To determine the degree to which the the unaveraged theory is invariant we 
need to calculate the variances of $\w$ as a function of $J.$ The smaller these variances the closer the threory is to being invariant. The variance of $\w $ is defined by,
\bea
\text{Var}\w &=& \big(\w^2\big) \cr \cr
&=& \int \w^2 \ P(J) dJ  
\label{(dotVarC)}
\eea
where we have used the fact that $\W =0.$


\subsection{Diagramatic Notation}
\begin{figure}[H]
		\begin{center}
			\includegraphics[scale=.9]{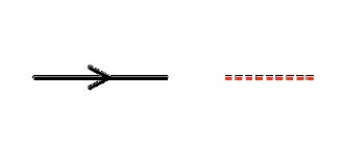}
			\caption{Fermion and $JJ$ propagators.}
			\label{cont}
		\end{center}
	\end{figure}

Let's introduce a diagramatic notation to represent the various steps in calculating $\text{Var}\w.$ Contractions of fermion operators will be represented as black solid directed lines, and contractions of $J$ when computing ensemble averages are represented as dashed red lines  (figure \ref{cont}).

\begin{figure}[H]
		\begin{center}
			\includegraphics[scale=.8]{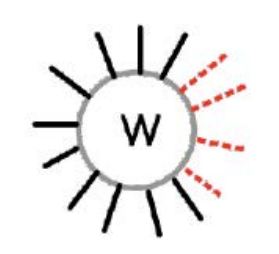}
			\caption{A diagnotic operator $W$ that depends on the fermion operators (black lines} and the couplings $J$ (dashed red lines).
			\label{w}
		\end{center}
	\end{figure}

	A diagnostic operator $W$ will depend on the fermionic variables $\psi, \bar{\psi}$ and also on the couplings $J.$ The $J$-dependence arises from the various factors of $e^{\pm i Ht}$ which can be expanded in powers of $H.$
		 Figure \ref{w} is a schematic representation for this dependence. A precise representation would involve a sum over different numbers of fermion operators and $J$ insertions.

\begin{figure}[H]
		\begin{center}
			\includegraphics[scale=.8]{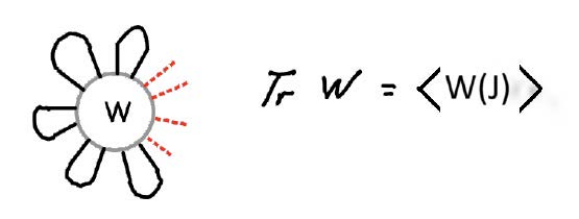}
			\caption{The expectation value of $W$ is computed by contracting the fermions but leaving the $JJ$ insertions uncontracted.}
			\label{exW}
		\end{center}
	\end{figure}

The expectation value of $W$ involves contracting the fermion lines in all possible ways, one of which is shown in figure \ref{exW}. The red dashed lines are uncontracted and the result is still a function of $\{J\}$.

\begin{figure}[H]
		\begin{center}
			\includegraphics[scale=.8]{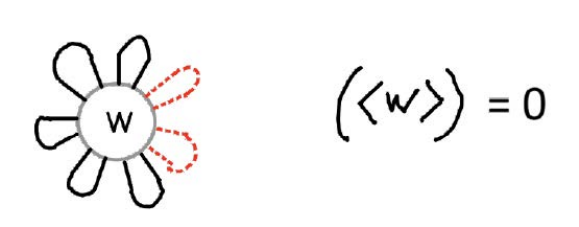}
			\caption{The final step in computing the average of the expectation value of $W$ is to contract the $JJ$ insertions.}
			\label{avexw}
		\end{center}
	\end{figure}

	Finally to calculate $\big(\langle W  \big\rangle \big)$ the red dashed lines must be contracted in all possible combinations (figure \ref{avexw}).   

	\bn The $SU(N)$ symmetry of the ensemble average insures that $\big(\langle W  \big\rangle \big) = 0.$

\begin{figure}[H]
		\begin{center}
			\includegraphics[scale=.8]{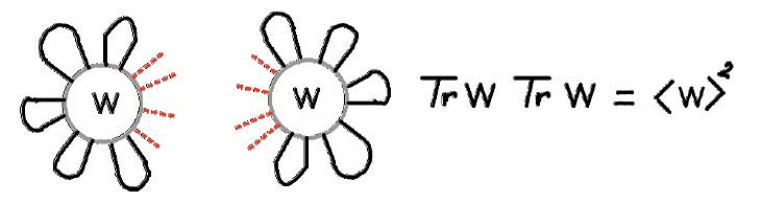}
			\caption{The expectation value of $WW$.}
			\label{exww}
		\end{center}
	\end{figure}

	To compute the variance we go through a similar series of steps. $\Tr W \  \Tr W$ (figure \ref{exww} ) is  obtained by juxtaposing two copies of figure \ref{exW}.
	%
	%
	Note that in figure \ref{exww} the fermion contractions do not bridge the gap from the left copy to the right copy. The red dashed lines are still uncontracted.

\begin{figure}[H]
		\begin{center}
			\includegraphics[scale=1.0]{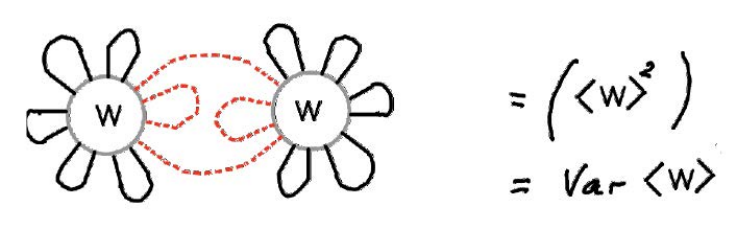}
			\caption{The average of the expectation value of $WW.$}
			\label{avexww}
		\end{center}
	\end{figure}

		In the final step, figure \ref{avexww}, the red dashed lines are contracted in all possible ways to give $\text{Var}   { \langle W \rangle  }.$

	The contractions in which no red line crosses the left-right gap sum to zero by virture of figure \ref{avexw}. The rest all have at least two red dashed lines crossing from left to right. These $JJ$ contractions are unaccompanied by black fermion lines.
	
	Now comes the main point: the unaccompanied red lines each give a factor (see equation \eqref{var})
	\bea  
&& \frac{N}{p^2} \  \frac{p! p!}{N^pN^p}  \cr \cr
  &\approx& \frac{N}{p^2}   \lf  \frac{p}{N} \rg^{2p} \cr \cr
  \eq  \frac{1}{\lambda} \lf  \frac{\lambda}{N}  \rg^{\sqrt{\lambda N}}
 \label{supress}
 	\eea
 	In the double-scaled limit, $N\to \infty$ with   
$\lambda=p^2/N$ fixed,
this goes to zero\footnote{%
The factors from the summation of fermion indices are not large 
enough to compensate for the small factor \eqref{supress}:
There are $p$ fermion loops for one $J$ in each copy of $W$, which
gives a factor $N^p/p!$; there are also $p$ fermion loops in the 
other copy of $W$, but they do not give another such factor, since 
the $JJ$ contraction between the two copies require both 
copies to have the same set of fermion indices.}
 faster than any power of $N.$
 	
 	To repeat and emphasize what we have found: 
 	In the ensemble average the symmetry-breaking-diagnostics all vanish as a consequence of the $SU(N)$ invariance of the ensemble averaged theory. But we have also found a much stronger fact---the  symmetry-breaking-diagnostics perfectly self-average. To put it simply, every measure of symmetry breaking vanishes faster than any power of $N$ in the double-scaled limit. This is particularly dramatic for the flat-space limit  in which $N$ is strictly infinite. In that limit we may assume that the $SU(N) $ global symmetry  of unaveraged  \dk \ is exact.  	
 	
 	The consequences are profound.
 	
\section{'t~Hooft Model  in The Bulk}

Exact global symmetries of a boundary theory---the boundary being the horizon in de Sitter holography---do not  translate into global symmetries of the bulk. Instead they are manifested as gauge symmetries  \cite{Banks:2010zn}.


This raises an interesting puzzle: Typically the entropy of a hot plasma
in  a theory with $SU(N)$ gauge symmetry is of order $N^2,$ the reason being that there are $N^2$ species of gauge bosons. On the other hand the maximum entropy in \dk \ is  only order $N.$ 
It would seem that there is a serious mismatch.

The resolution of this puzzle is simple---there are no gauge bosons in $(1+1)$ dimensions;  gauge fields in $(1+1)$ dimensions are non-dynamical. Their only role is to mediate instantaneous linear potentials between charges, but they are not independent degrees of freedom. The entropy all comes from matter fields.
The entropy of a hot $QCD_2$  plasma\footnote{We use $QCD_2$ as shorthand for $(1+1)$ dimensional QCD.} is due to the quark fields and  is  $\sim N.$ Similarly in \dk \ the entopy is due to the $N$ species of fermions; it is also $\sim N.$ In both cases the fermions  transform as fundamentals under $SU(N).$ It is completely reasonable and even compelling to identify the \dk  \ fermions with $QCD_2$  quarks. 

These  considerations lead to the conclusion that 
 the bulk dual of \dk \ in the flat-space limit is a $(1+1)$ dimensional $SU(N) \times U(1)$ gauge theory, with a single multiplet of fermions in the fundamental representation. Moreover the flat-space limit requires that we take the $N\to \infty $   limit.   Simply put,  the bulk dual  is the one-flavor 't~Hooft model\footnote{The $N$ species of SYK fermions are often referred to as flavors but it is more correct to think of them as colors.} 	 \cite{tHooft:1974pnl}. For brevity we'll refer to this as $QCD_2.$
 
 In this section we will discuss and compare the dimensionless parameters that occur in \dk \ and in its dual---one-flavor $QCD_2.$   In both cases there is only one dimensionless parameter and the duality requires a correspondence between them.

\subsection{The 't~Hooft and Flat Space Limits}

The standard large $N$ limit of gauge theories---say in four dimensions---is the 't~Hooft limit in which the dimensionless gauge coupling $\g$ goes to zero, with the dimensionless  't~Hooft coupling\footnote{The 't~Hooft coupling $\g^2 N$ is usually denoted by $\lambda.$ To avoid confusion with the \dk \ parameter $\lambda$ we denote the 't~Hooft coupling by $\alpha.$}
 $\alpha$ remaining finite.
\be 
\alpha \equiv \g^2 N  \  (\alpha \  \text{finite in the 't~Hooft limit})
\label{tooftlim}
\ee
The flat space  limit   is different; it is a super-strongly coupled  limit in which the
 gauge  coupling $\g$ is kept fixed and $\alpha \to \infty$. This was discussed at length in  \cite{Sekino:2025bsc}.

Something similar is true for $QCD_2$ but, because $\g$ and $\alpha$ are not dimensionless, extra care is needed in defining the limits\footnote{%
We thank Ofer Aharony for helpful comments on this point.}.  
We will need to construct dimensionless versions of them in order to define the limits correctly.

In  $QCD_2$  there are two DIMENSIONFUL parameters, the quark mass\footnote{We use the notation $m_q$ for the shifted (renormalized) quark mass defined in terms of the Lagrangian mass $m_0$ by $m_q^2 = m_0^2 -\g^2N/\pi.$}   $m_q$ and the gauge coupling $\g.$ Both have units of mass. We define the dimensionless coupling parameter $\bar{g}^2$ to be the ratio 
\footnote{This definition of $\bar{g}^2$ is different from the one in the previous versions of this paper. Previously, we have defined a barred (dimensionless) quantity by dividing the original quantity by $\mu = m_{q}^2/N$, while in the present version we define it by dividing by $m_{q}^2$.} 
\be
\bar{g}^2 = \frac{\g^2}{m_q^2} ~ . \label{defgbar}
\ee
The bar notation in \eqref{defgbar} is used to represent the dimensionless version of a quantity obtained by dividing by an appropriate power of $m_{q}.$  

Similarly the  't~Hooft coupling  $\alpha = \g^2 N$ is not dimensionless; its dimensionless version is\footnote{%
Again, this definition of $\bar{\alpha}$ differs from that in the previous versions of this paper.},
\be  
\bar{\alpha} =\frac{\alpha}{m_{q}^2} = \bar{g}^2  N.
\label{baralph}
\ee

The 't~Hooft Hamiltonian, which describes mesons in the light cone frame,
is the sum of the free quark term and potential term,
\be 
H = H_0 +V.   
\label{Htot}
\ee
The free quark term is
\be  
H_0 = \frac{m_q^2 }{x(1-x)} \ \ \ \ (0<x<1),
\label{H0}
\ee
where $x$ is the ``longitudinal fraction" of momentum.
 
The potential term results from a single gluon exchange between 
a quark $q$ and antiquark $\bar{q}$, and is given by 
\be
V =\g^2 N  |r|,
\label{V}
\ee
where $|r|$ is the distance between $q$ and $\bar{q}$. 
The quantity
\be  
\tau \equiv \g^2 N=\alpha
\label{tau}
\ee
is called the string tension. It---not $\g^2$---is the measure of interaction strength. 

The full Hamiltonian has the form,
\be  
H=m_q^2\lf \frac{1}{x(1-x)} +{\bar\alpha}|r| \rg.
\ee
The parameter $\bar\alpha$ as defined in \eqref{baralph}
is the dimenionless coupling which characterizes the 
interaction. It is usually kept finite 
in the $N\to \infty$ limit~\cite{tHooft:1974pnl}.

We may now define the flat-space limit and the 't~Hooft limit in terms of $\bar{g}$ and $\bar{\alpha}$ as follows.

\bn

\bn
Flat-space limit:   $N\to \infty, \ \ (\bar{g} \ \ \text{fixed}, \ \ \bar{\alpha} \to \infty).$
\bn

\bn
't~Hooft limit:  \ \  \  $N\to \infty, \ \ (\bar{\alpha} \ \ \text{fixed}).$
\bn

To understand how the limits work we will follow  \cite{Sekino:2025bsc}	 adapted to the two-dimensional case.
Consider the form of the $1/N$ expansion for some dimensionless amplitude. It has the form,
\be 
\CA=\sum_{n=0}^{\infty}\frac{\tilde{\mathcal{F}}^{(n)}(\bar{\alpha},L)}{N^n}
\label{A}
\ee
where each 
$\tilde{\mathcal{F}}^{(n)} (\bar{\alpha},L)$
is given by
the sum of diagrams with genus\footnote{In pure gauge theory the genus of any diagram is an integer. Quarks introduce boundaries, for example a planar diagram with one quark loop has the topology of a disk which has genus $1/2.$} 
$n/2$. 
Each diagram is IR divergent as in the real QCD, so
we need an IR cutoff length $L$ in 
the expansion \eqref{A}. 
We define $L$ to be dimensionless, by measuring it
relative to $1/\sqrt{\alpha}$.

The functions
$\tilde{\mathcal{F}}^{(n)} (\bar{\alpha}, L)$ 
take the form,
\bea
\tilde{\mathcal{F}}^{(n)} (\bar{\alpha},L)
  & = &  \bar{\alpha}^n \times \mathcal{F}^{(n)} (\bar{\alpha}, L) 
\label{tildeF}
\eea
with each $\mathcal{F}^{(n)} (\bar{\alpha}, L)$ 
being a power series in $\bar{\alpha}$ 
starting from the zeroth power of $\bar{\alpha}$.
(It is a power series in 
$\bar{\alpha}L^2$ for large $L$.) 
The form \eqref{tildeF} reflects the fact that an order-$n$ diagram has at least $2n$ vertices (i.e., $n$ factors of $\alpha$)~\cite{Sekino:2025bsc}, and that the power of the cutoff $L$ for the leading diagram for each $n$ does not depend on $n$. 
An example of this is to increase $n$
by one from a given diagram, by
 inserting a quark loop into a free gluon propagator: The power of $\alpha$ increases by one, but the infrared behavior of the corrected gluon propagator does not change from the free propagator, as dictated by gauge invariance, so no extra dependence on $L$ is introduced by this operation. 

If the cutoff length is sufficiently larger than the dynamical scale, the amplitude should be independent of the cutoff. 
We assume that the cutoff dependence disappears from the final result if $L\gg 1$, i.e., if the cutoff length is much larger than the string scale $1/\sqrt{\alpha}$.
This means that after summing over the infinite series for each $n$, the cutoff $L$ 
can be effectively replaced with an order 1 number (which will be taken to be 1 below for simplicity); for $L\gg 1$, we will have
\be
\mathcal{F}^{(n)} (\bar{\alpha}, L)\sim
\mathcal{F}^{(n)} (\bar{\alpha}).
\label{Lsim1}
\ee
Substituiting \eqref{tildeF} and \eqref{Lsim1} into \eqref{A} gives, 
\bea 
\CA &= &\sum_{n=0}^{\infty}  \left( \frac{\bar{\alpha}}{N} \right)^n \mathcal{F}^{(n)} (\bar{\alpha}) \cr\cr
 & = &\sum_{n=0}^{\infty}  \bar{g}^{2n} \mathcal{F}^{(n)} (\bar{\alpha}),
\eea
assuming the cutoff length is sufficiently large.

The flat-space limit is obtained by letting $\bar{\alpha}\to \infty$ with $\bar{g}$ kept finite. 
For this limit to make sense---we assume that it does---
the functions 
$\mathcal{F}^{(n)} (\bar{\alpha})$ must have finite limits \cite{Sekino:2025bsc},
\be 
\lim_{\bar{\alpha} \to \infty} \mathcal{F}^{(n)} (\bar{\alpha})= \CF_n.
\label{lims}
\ee
The amplitude in the flat-space limit  is given by,
\be 
\CA=\sum_{n=0}^{\infty} \CF_n    \bar{g}^{2n}.
\label{Ainfty}
\ee

Equation \eqref{Ainfty} has the following interpretation: The sum is over worldsheets of genus $n/2.$ The factors $\CF_n$ are the worldsheet amplitudes for genus $n/2.$ And most interestingly the factor $\bar{g}^{2n}$
is the dependence on the open string coupling constant $g_{st}$ for a worldsheet of genus $n/2$, with the identification,
\be
g_{st}=\bar{g}.
\label{gst}
\ee

We note that the only dimensionless parameter in the flat space limit of $QCD_2$\ is the open string coupling  $g_{st}.$

\subsection{DSSYK-QCD$_2$ Correspondence}

\dk \ in the flat-space limit also has a single dimensionless paramter,  $\lambda.$ If there is a correspondence relating \dk \ and $QCD_2$ then $\bar{g}$ and $\lambda$ must be related.

\dk \ also has a $1/N$ expansion with essentially the same form  as \eqref{A} but with 
$\bar{\alpha}$ being replaced by $p^2$~\cite{Sekino:2025bsc}
\be 
\CA=\sum_n\frac{\tilde{\mathcal{F}}^{(n)}(p^2)}{N^n}.
\label{Aexp}
\ee
The duality between \dk \ and $QCD_2$ requires that we make the identification\footnote{This identification of the parameters is different from the one 
adopted in the previous paper by two of the present authors~\cite{Sekino:2025bsc}, in which the similarity of perturbation expansions in SYK and QCD has been discussed. In that paper, QCD with degrees of freedom in the adjoint representation was considered, and the authors found the identification $p=\alpha$.
In that case, $N$ in SYK corresponds to $N_{ym}^2$ in QCD, as naturally expected from the matching of the number of the degrees of freedom, which leads to the identification $\lambda=g_{ym}^4$.} 
\begin{equation}
p^2 = \bar{\alpha}.
\end{equation}

Dividing by $N$ and recalling that $\frac{p^2}{N} =\lambda,$ we find the correspondence,
\be 
\lambda = \bar{g}^2=g_{st}^2.
\label{L=g2}
\ee
Thus $\lambda$ corresponds to the square of the open string coupling.


\bn\\

\subsection{The Meson Tower} \label{mesontower}
\begin{figure}[H]
		\begin{center}
			\includegraphics[scale=.8]{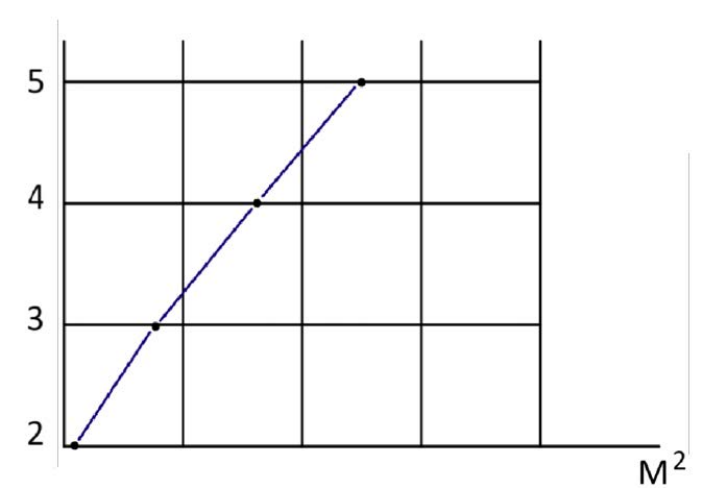}
			\caption{The meson Regge   trajectory for a particular value of $\mu$.}
			\label{tower}
		\end{center}
	\end{figure}

The 't~Hooft model is  the  large $N$ limit of a gauge theory, but it may also be thought of as an open string theory.
The spectrum of mesons in the  model consists of a single tower or Regge-like trajectory.
 Figure \ref{tower} is taken from  \cite{tHooft:1974pnl} and shows the Regge trajectory\footnote{%
See \cite{Fateev:2009jf, Litvinov:2024riz, Artemev:2025cev} for recent work on the mass spectrum in the 't~Hooft model. We thank Pavel Meshcheriakov for discussion on these results.} for a particular non-zero value of $m_{q}.$ The horizontal axis is the meson mass squared; the vertical axis is an integer  $n\geq 2$  analogous to angular momentum in higher dimensions. 
Asymptotically for large $n$ the trajectory approaches a straight line.
The states along the tower can be identified with a set of operators,
\be 
T_n =\bar{q}_i \  \frac{d^n q^i}{dt^n} \ \ \ \ \ \ \ \text{sum on $i$  implicit}
\label{qq}
\ee
where $q$ represents quark fields. For $n=0$ and $n=1$ the operators \eqref{qq} are trivial, namely they are the conserved $U(1)$ charge and the Hamiltonian itself. As such they are  have no time dependence. The dynamical meson states begin at $n=2.$
The meson states alternate in their time reversal properties. For even $n$ they are time-reversal odd; for odd $n$ they are time-reversal even. 

The operators in \eqref{qq} are in a sense complete in that any gauge invariant operator is a function of them. In addition to the single meson states there are multiple meson states and these span the space of states of the 't~Hooft model.

Is there a \dk \ dual to the meson tower?
Consider singlet operators studied in \cite{Gross:2017aos}, 
\be 
S_n =\bar{\psi}_i \  \frac{d^n \psi^i}{dt^n}.
\label{grros}
\ee
These operators also form a tower which (we believe) is complete in the sense that any singlet operator is a function of the set ${S_n}.$ For example products $S_nS_m$ are singlets.
At the moment we don't know the masses of the states created by the operators in \eqref{grros}. One clear prediction is that the spectrum of the \dk \ singlets for small $\lambda$ should match that of the 't~Hooft model. 

One thing that we do know is that in both cases the time-reversal properties along the tower alternate between odd and even with the lowest state being odd under time-reversal.  

We will come back to this in a future publication.

\section{QCD in Rindler Space and DSSYK$_{\infty}$}   \label{S:qcdr }

\dk \ is conjectured to be a holographic description of the static patch of JT-de Sitter space. In the flat-space limit the static patch becomes ordinary $(1+1)$-dimensional  Rindler space. In \cite{Susskind:2023hnj}\cite{Sekino:2025bsc}   we discussed QCD in Rindler space. We  review that discussion here.

	 It should be possible to formulate any quantum field theory in Rindler space but, unlike the case of the light cone frame, there is not a great deal of research on QFT in Rindler coordinates. We will  fill the gap with some
intuitive (but surely correct) observations.

\subsection{The Phase Boundary and the Stretched 
Horizon}\label{S:pbsh }

For the moment we will not specify the dimension $D$ of spacetime. 

The metric of Rindler space is,
\be 
ds^2 = -\rho^2 dt^2 +d\rho^2 +dx^idx^i
\label{Rmtric}
\ee
where $x^i$ are coordinates parameterizing the  $(D-2)$-dimensional plane of the horizon.  

\begin{figure}[H]
\begin{center}
\includegraphics[scale=0.3]{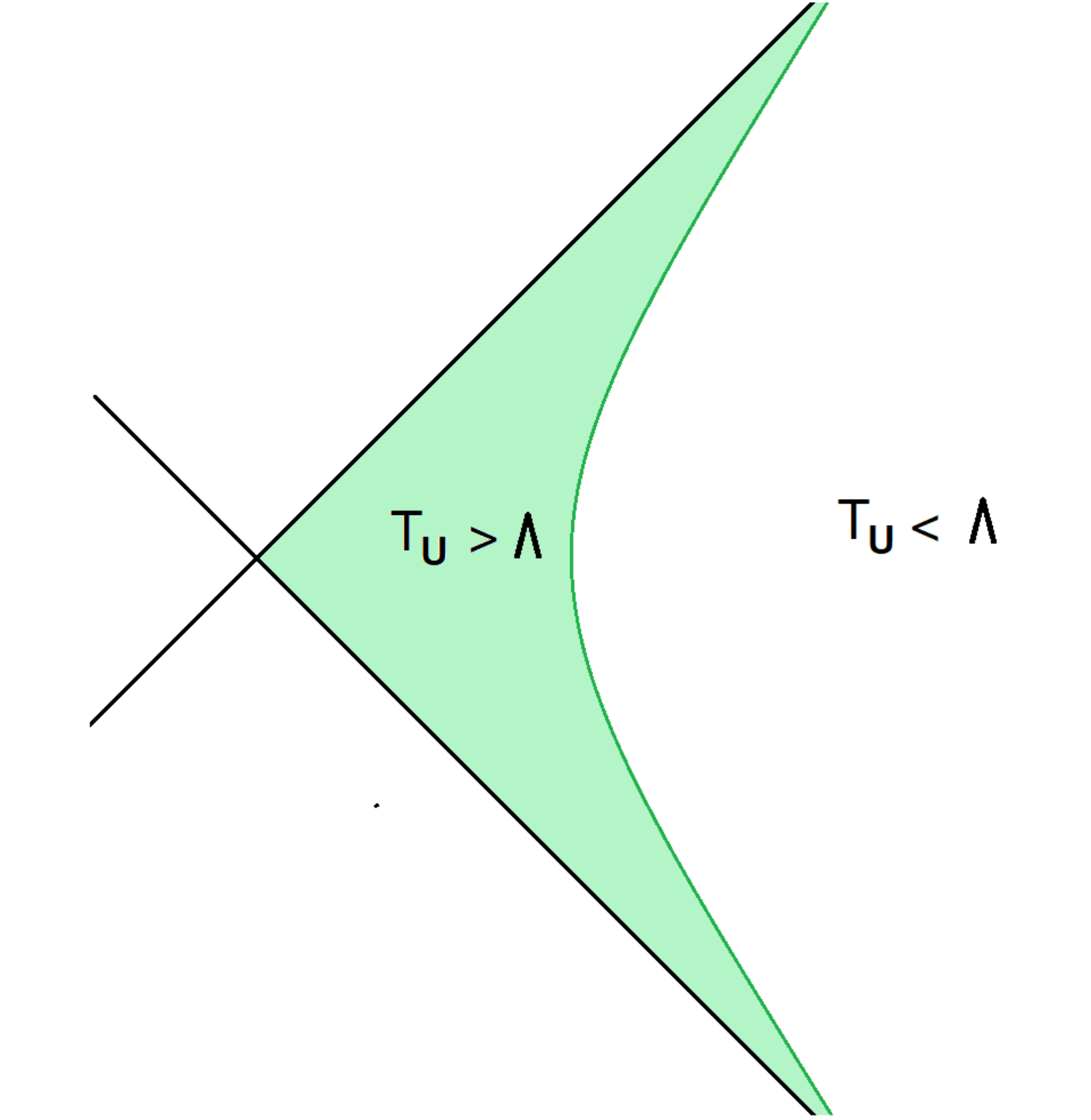}
\caption{Rindler space divided into hot and cold regions.}
\label{rindler}
\end{center}
\end{figure}

The vacuum in Rindler space is described as a thermal state with dimensionless temperature $T_R = \frac{1}{2\pi}.$ The proper temperature registered by a thermometer located at distance $\rho$ from the horizon is
\be 
T_U(\rho) = \frac{1}{2\pi \rho}.
\label{unruh}
\ee
Let us introduce a mathematical $(D-1)$-dimensional time-like surface of fixed $\rho$ at the value of $\rho$ for which 
\be
T_U(\rho) = \Lambda
\label{Tu=Labda}
\ee
where $\Lambda$ is the  QCD energy scale. To give it a precise definition $\Lambda$ can be taken to be the  temperature of the QCD confinement--de-confinement transition.  The surface (shown in figure \ref{rindler}) separates Rindler space into two regions: a hot plasma region where  QCD is in the deconfined phase  and a cold region 
 where it is in the confined phase. This is shown in figure \ref{rindler}.
We may think of the surface as the phase boundary between the unconfined QCD-plasma phase and the confined phase. Alternatively it is the stretched horizon.

\bn
\it   The QCD phase boundary is  the stretched horizon. \rm

\bn
 It is a string scale distance from  the mathematical horizon at $\rho =0.$

For $D>2$ in the hot deconfined region quarks and gluons propagate freely and the entropy per unit area is of order $N^2.$ For $D=2$ there are no dynamical gluons; the entropy is order $N,$ the number of quark species.
In the confined region only $SU(N)$ singlets, i.e.,  mesons, and for $D>2$ glueballs, can propagate
\footnote{Baryons have mass proportional to $N$ and in the $N\to \infty$ disappear from the spectrum.}.
The number of species of hadrons with mass less than or order $\Lambda$ is finite and independent of $N.$ Therefore the entropy per unit area in the outer confined region is order one. 

To a high approximation the phase boundary is the end of the world. The outer confined region is almost completely empty; the vast majority of degrees of freedom cannot escape the hot region. Only the light $SU(N)$ singlets can escape. When a quark  from the hot region hits the phase boundary it is reflected back with a probability very close to $1.$ With a small probability
 it passes through the boundary dragging an antiquark with it, the two forming a meson. A similar thing can be said for gluons.
In other words to quote   \cite{Susskind:2023hnj} ``almost everything is confined" (to the hot plasma-like stretched horizon)\footnote{The word confined has two different meanings in this context. Non-singlet degrees of freedom are \it confined \rm to the deconfined (green) region. In the confined region quarks and gluons are \it confined \rm to form singlet hadrons.}.

 \subsection{Escape from the Stretched Horizon}\label{escape}

It was explained in \cite{Susskind:2023hnj} that there are far too many degrees of freedom in any holographic description of de Sitter Space for all, or even a tiny fraction, to propagate into the interior of the static patch. Almost all the degrees of freedom comprising the entropy must be confined to the immediate  vicinity of the horizon.  This is one sense in which we use the word confined---confined to the stretched horizon.

\begin{figure}[H]
\begin{center}
\includegraphics[scale=.8]{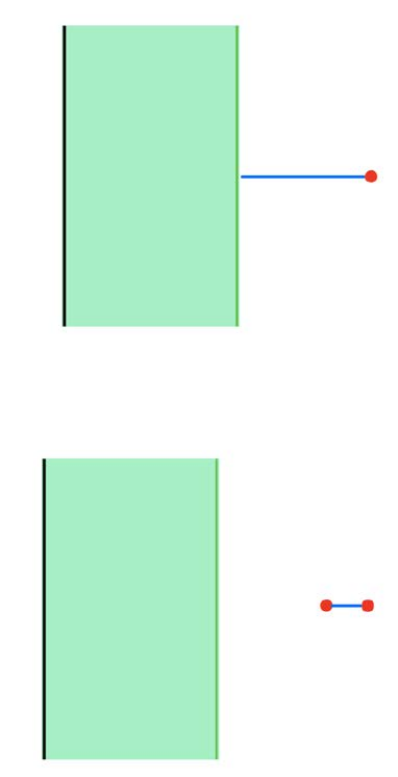}
\caption{A quark which leaves the hot region is connected to the plasma by a QCD string. The quark can only escape if the string breaks resulting in a singlet meson emitted into the cold region.}
\label{confine}
\end{center}
\end{figure}

All of this is well understood in the
 example of QCD in Rindler space where the mechanism is ordinary confinement. In the hot plasma-like  stretched horizon quarks  are free to propagate independently but when they try to escape into the cold region they are held in place by QCD strings illustrated in the schematic figure \ref{confine}.
The objects which can propagate away from the QCD plasma into the static patch are the singlet, mesons and glueballs for $D>2$  (just mesons for $D=2.$). This is the second way we use the term confined: in the cold bulk region quarks are confined into hadrons.

 The spectrum of singlets is very sparse and unlike the spectrum of quarks (or quarks and gluons) it  does not grow with increasing $N.$ This resolves a certain tension noted earlier in \cite{Susskind:2023hnj}. In the present $D=2$ case the total number of degrees of freedom is $N$ as seen from the fact that the entropy of \dk \ is exactly $$N\log{2},$$ but the number of degrees of freedom in the bulk is much less, indeed the entropy found in the static patch for $\rho > \ell_{string} $ is order unity. The resolution of this tension is the two types of confinement---confinement of $SU(N)$ charge to the plasma-like stretched horizon, and confinement to singlets in the cold bulk. 
 
 \section{Conclusions}
A  holographic duality  should be  more than just a claim of a bulk geometry; it should also determine  the matter fields and particles  which can propagate in the bulk as well as the interactions. Ideally a holographic description of our own universe would include the standard model, neutrino masses, and other features of particle physics. We have taken up the challenge of determining the bulk theory for a particular example: the holographic duality between double-scaled SYK and  JT-de Sitter space. The conjecture that \dk \ is dual to JT-de Sitter is not new but the identification of the bulk matter sector is. We've given evidence  that the bulk theory is the ultra-strongly coupled fixed $\bar{g}$ limit of  $(1+1)$ dimensional QCD---a string theory with mesons lying on a single Regge trajectory.  Unlike the limit 't~Hooft limit with $\bar{\alpha}$ held fixed, the fixed $\bar{g}$ limit  is a theory of interacting meson-strings with $g_{st}=\bar{g}$ being the open string coupling constant. On the \dk \ side the role of 
$g_{st}^2$ is played by $\lambda.$ 
The limit $\lambda \to 0$ corresponds to the standard 't~Hooft limit, in which mesons are non-interacting.

The  $QCD_2 $ gauge forces  result from an emergent $SU(N)$ symmetry of \dk \ which becomes perfectly self-averaging in the double-scaled limit. The quarks of the model are none other than the SYK fermions. The gauge forces lead to two related concepts of confinement. The first is conventional; only gauge singlets propagate in the bulk. The second is that all the non-singlet \dof  including the fermions themselves are confined to the stretched horizon, where they account for the large entropy  of de Sitter space. This is a very interesting fact---the same theory that accounts for the relatively sparse spectrum in the bulk also accounts for the huge  entropy in the stretched horizon.

The correctness of this duality would be no small thing. It would be the first example of a theory with ``sub-cosmic locality"  not dependent on supersymmetry to maintain the locality. It would also seem to contradict the claim that without supersymmetry a  small cosmological constant requires fine-tuning, as well as the claim that quantum de Sitter space is at best metastable. The importance of testing the lore that has grown up around de Sitter space, by means of concrete examples, cannot be overstated. 

Finally a comment about \dk \ in the Majorana case with real fermions: In that case the $SU(N)$ symmetry is replaced by $O(N)$ and the bulk gauge theory is an $O(N)$ gauge theory with a single multiplet of fermions in the fundamental. The main difference is that half of the states along the tower are removed; only the time-reversal even states survive. The same is true of the tower of SYK-singlet operators.

\section*{Acknowledgement}
We would like to thank Juan Maldacena and Ofer Aharony
for valuable comments. We also thank Henry Lin, Steve
Shenker, Douglas Stanford for discussions. Part of this
work has been done while Y.S. was visiting Stanford
Institute for Theoretical Physics (SITP) on sabbatical
leave from Takushoku University under the ``Long-term Overseas Research'' program. He is grateful to Takushoku
University for support and SITP for hospitality. Part of
this work has been done while S.M. was visiting
SITP under the ``Oversea Visiting Program for Young
Researchers'' funded by the Extreme Universe (ExU)
collaboration. The work of Y.S. and S.M. is supported
in part by MEXT KAKENHI Grant Number 21H05187.
The work of S.M. is also supported in part by the National Science and Technology Council (No. 111-2112-
M259-016-MY3).

	\end{document}